\documentclass[a4paper,11pt]{article}
\usepackage{pos}
\usepackage{color}

\newcommand{\ba}{\begin{array}}
\newcommand{\ea}{\end{array}}

\newcommand{\Dslash}{\relax{\kern+.25em / \kern-.70em D}}

\newcommand{\GeV}{{\rm GeV}}

\newcommand{\Real}{\relax{\mathsf{\Gamma\kern-.35em R}}}
\newcommand{\Int}{\relax{\mathsf{Z\kern-.40em Z}}}

\newcommand{\NF}{N_\mathrm{\scriptstyle f}}

\newcommand{\SF}{{\rm SF}}

\newcommand{\gbar}{\kern1pt\overline{\kern-1pt g\kern-0pt}\kern1pt}
\newcommand{\mbar}{\kern2pt\overline{\kern-1pt m\kern-1pt}\kern1pt}
\newcommand{\obar}[1]{\kern3pt\overline{\kern-2pt #1\kern-0pt}\kern1pt}

\newcommand{\lQCD}{\Lambda_{\rm\scriptscriptstyle QCD}}
\newcommand{\MW}{M_{\rm\scriptscriptstyle W}}

\newcommand{\hopc}{\kappa_{\rm c}}

\newcommand{\fP}{f_{\rm\scriptscriptstyle P}}

\newcommand{\fA}{f_{\rm\scriptscriptstyle A}}

\newcommand{\gP}{g_{\rm\scriptscriptstyle P}}

\newcommand{\gA}{g_{\rm\scriptscriptstyle A}}

\newcommand{\ZP}{Z_{\rm\scriptscriptstyle P}}

\newcommand{\zf}{z_{\rm f}}
\newcommand{\ZPSF}{Z_{\rm\scriptscriptstyle P}^{\rm\scriptscriptstyle SF}}
\newcommand{\ZPchiSF}{Z_{\rm\scriptscriptstyle P}^{\rm\scriptscriptstyle \chi SF}}

\newcommand{\sigmaP}{\sigma_{\rm\scriptscriptstyle P}}

\newcommand{\SigmaP}{\Sigma_{\rm\scriptscriptstyle P}}

\newcommand{\SigmaPSF}{\Sigma_{\rm\scriptscriptstyle P}^{\rm\scriptscriptstyle SF}}
\newcommand{\SigmaPchiSF}{\Sigma_{\rm\scriptscriptstyle P}^{\rm\scriptscriptstyle \chi SF}}

\newcommand{\icA}{c_{\rm\scriptscriptstyle A}}

\newcommand{\uSF}{u_{\rm\scriptscriptstyle SF}}

\newcommand{\abar}{\kern1pt\overline{\kern-1pt a\kern-0.5pt}\kern1pt}

\title{Quark mass RG-running for $N_f$ =3 QCD in a $\chi SF$ setup}

\author[a]{\begin{center}
     \includegraphics[height=2.0\baselineskip]{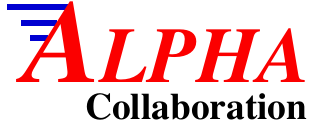}
\end{center} Isabel Campos Plasencia}
\author[b]{Mattia Dalla Brida}
\author[c,d]{Giulia Maria de Divitiis}
\author[e]{Andrew Lytle}
\author[f,g]{Mauro Papinutto}
\author*[c,d]{Ludovica Pirelli}
\author[d]{Anastassios Vladikas}

\affiliation[a]{Instituto de F\'isica de Cantabria IFCA-CSIC,\\
  Avda. de los Castros s/n, 39005, Santander, Spain}
 \affiliation[b]{Theoretical Physics Department, CERN, \\
 CH-1211, Geneva 23, Switzerland}
 \affiliation[c]{Dipartimento di Fisica, Universit\`a di Roma ``Tor Vergata'', \\
 Via della Ricerca Scientifica 1, 00133 Roma, Italy}
 \affiliation[d]{INFN, Sezione di Tor Vergata, \\
 Via della Ricerca Scientifica 1, 00133 Roma, Italy}
 \affiliation[e]{Department of Physics, University of Illinois at Urbana-Champaign, \\
 Urbana, Illinois, 61801, USA}
 \affiliation[f]{Dipartimento di Fisica,  Universit\`a di Roma La Sapienza, \\
 Piazzale A.~Moro 2, 00185 Roma, Italy}
\affiliation[g]{INFN, Sezione di Roma, \\
Piazzale A.~Moro 2, 00185 Roma, Italy}

\emailAdd{isabel.campos@csic.es}
\emailAdd{mattia.dalla.brida@cern.ch}
\emailAdd{giulia.dedivitiis@roma2.infn.it}
\emailAdd{atlytle@illinois.edu}
\emailAdd{mauro.papinutto@roma1.infn.it}
\emailAdd{ludovica.pirelli@roma2.infn.it}
\emailAdd{tassos.vladikas@roma2.infn.it}

\abstract{We compute the nonperturbative quark mass RG-running in the range $\lQCD\lessapprox\mu\lessapprox\MW$ for $\NF=3$ massless QCD with a mixed action approach: sea quarks are regularised using nonperturbatively $O(a)$-improved Wilson fermions with Schr\"odinger functional (SF) boundary conditions, employing the configurations of ~\cite{Campos:2018ahf}, while valence quarks are regularised using nonperturbatively $O(a)$-improved Wilson fermions with chirally rotated Schr\"odinger functional boundary conditions ($\chi$SF). Our result is compatible with its SF counterpart of ref.~\cite{Campos:2018ahf}, confirming the universality of $\chi$SF and SF in the continuum limit. We also establish the optimal tuning strategy for the critical hopping parameter $\hopc$ and the $\chi$SF boundary counterterm coefficient $\zf$.
We work in two energy regimes with two different definitions of the coupling: SF-coupling for 2 \GeV~$\lessapprox\mu\lessapprox\MW$ and GF-coupling for $\lQCD \lessapprox\mu\lessapprox 2~\GeV$.}

\FullConference{%
 The 38th International Symposium on Lattice Field Theory, LATTICE2021
  26th-30th July, 2021
  Zoom/Gather@Massachusetts Institute of Technology
}

 \tableofcontents

\begin{document}
\begin{flushright}
CERN-TH-2021-205
\end{flushright}
\maketitle

\section{Introduction}
We compute the quark mass RG-running in the range $\lQCD\lessapprox\mu\lessapprox \MW$ for $\NF=3$ lattice theory with Wilson-clover fermions obeying chirally rotated Schr\"odinger functional ($\chi$SF) boundary conditions, which is a variant of the Schr\"odinger functional (SF) renormalisation scheme. The special feature of this choice is that continuum massless QCD with $\chi$SF  boundary conditions is equivalent to the one with SF boundaries, as one is obtained from the other by suitable redefinitions of the fermion fields; yet, on the lattice, in the $\chi$SF we obtain automatically $O(a)$-improved renormalisation parameters and lattice step scaling functions. ~\cite{Sint:2010eh, Sint:2010xy, Mainar:2016uwb}\\
The possibility of automatic $O(a)$-improvement is the main reason to adopt $\chi$SF in a long-term project, outlined in ~\cite{Campos:2019nus}, which ultimately aims at providing the step scaling matrices of all four-fermion operators that contribute to $B_{\rm\scriptscriptstyle{K}}$ in the Standard Model and beyond. Therefore, it is first necessary to perform tests on $\chi$SF and make comparisons with SF results, in analogy to those performed in ref. \cite{DallaBrida:2016smt}. In the present talk we compare a $\chi$SF estimate of $M/\mbar(\mu)$ to the SF one in the same energy range and analogous setup, cf. ref.~\cite{Campos:2018ahf}. $M$ is the RGI quark mass, and $\mbar(\mu)$ is the renormalised quark mass at energy scale $\mu$ in the SF or $\chi$SF scheme (they are the same scheme in the continuum). The next step of our project- the computation of the tensor operator- is outlined in ref. \cite{Plasencia:2021ihd}.

\section{Definitions in SF and in $\chi$SF}
At a formal level, continuum massless QCD with $\chi$SF boundary conditions is obtained from its SF counterpart by a chiral non-singlet transformation of the fermion fields ~\cite{Sint:2010eh}: 
\begin{equation}
\label{eq:ferm-rots}
\psi = R(\pi/2) \, \psi^\prime \,\, , \qquad  \bar \psi = \bar \psi^\prime \,  R(\pi/2) \,\, ,
\end{equation}
where  $\psi, \bar \psi$ and $\psi^\prime, \bar \psi^\prime$ are doublets in isospin space and $R(\alpha) = \exp(i \alpha \gamma_5 \tau^3/2)$. We can map SF correlation functions into $\chi$SF ones ~\cite{DallaBrida:2016smt}. For example, the boundary-to-bulk correlation functions
\begin{equation}
\label{eq:bound_to_bulk}
\begin{array}{ccc}
\fP\equiv -\dfrac{1}{2} \langle P^{ud} {\cal O}_5^{du} \rangle_{(\SF)} &\, ,& \gP^{ud} \equiv -\dfrac{1}{2} \langle P^{ud} {\cal Q}_5^{du} \rangle_{(\chi \SF)} \, ,
\end{array}
\end{equation}
and the boundary-to-boundary correlation functions
\begin{equation}
\label{eq:bulk_to_bulk}
\begin{array}{ccc}
f_1 \equiv -\dfrac{1}{2}  \langle {\cal O}_5^{ud} {\cal O}_5^{\prime du} \rangle_{(\SF)} &\, ,& g_1^ {ud} \equiv -\dfrac{1}{2} \langle {\cal Q}_5^{ud} {\cal Q}_5^{\prime du} \rangle_{(\chi \SF)}   \ ,
\end{array}
\end{equation}
are related and satisfy ~$\fP \, =\, \gP^{ud}$, $f_1 = g_1^{ud}$. Notation is standard: $u,d$ are flavour indices; $P^{ud}$ is the non-singlet pseudoscalar density; ${\cal O}_5, {\cal O}_5^{\prime}$ and ${\cal Q}_5, {\cal Q}_5^{\prime}$ are respectively SF and $\chi$SF boundary operators. The latters are defined in ref.~\cite{DallaBrida:2016smt}. 
The above formal properties follow from the invariance of the massless QCD action under flavour and chiral transformations. They are broken on the lattice, but  they are recovered after renormalisation in the continuum limit.
We can define renormalisation conditions in SF and in $\chi$SF setups for the pseudoscalar operator,
\begin{equation}
\label{eq:renorm_cond}
\begin{array}{ccc}
\dfrac{\ZPSF(g_0^2,L/a) \fP(T/2)}{\sqrt{f_1}}= \Bigg [ \dfrac{\fP(T/2)}{\sqrt{f_1}} \Bigg ]_{g_0^2=0}&\, ,&
\dfrac{\ZPchiSF(g_0^2,L/a) \gP^{ud}(T/2)}{\sqrt{g_1^{ud}}}=\Bigg [ \dfrac{\gP^{ud}(T/2)}{\sqrt{g_1^{ud}}} \Bigg ]_{g_0^2=0} \, ,
\end{array}
\end{equation}
for a symmetric lattice with volume $L^3 \times T$  and for $T=L$.
The above relations evince that the renormalisation scale is $\mu = 1/L$.
The definition of the step scaling functions for the pseudoscalar operator immediately follows
\begin{equation}
\label{eq:sigma_def}
\Sigma_{\rm P}^{SF,\chi SF}(g_0^2,a/L)=\dfrac{Z_{\rm P}^{SF,\chi SF}(g_0^2,2L/a)}{Z_{\rm P}^{SF,\chi SF}(g_0^2,L/a)} \, ,
\end{equation}
and the relations between SF and $\chi$SF imply that $\SigmaPSF(g_{0}^{2},a/L)$  and $\SigmaPchiSF(g_{0}^{2},a/L)$ have the same continuum limit $\sigmaP(u)$
\begin{equation}
\label{eq:sigma_lim}
\sigmaP(u)=\underset{a\rightarrow0}{lim} \ \Sigma_{\rm P}^{SF,\chi SF}(g_{0}^{2},a/L)\Big \vert_{\gbar^2(L) = u} \,\ ,
\end{equation}
where $u$ is the squared renormalised coupling.

\section{Computational setup}
We use the configuration ensembles of ~\cite{Campos:2018ahf}, thus working in a mixed action setup: sea quarks are regularised in SF, but we invert the Dirac-Wilson operator with $\chi$SF boundary conditions. 
The ensembles span over two energy regimes with an intermediate ("switching") scale conventionally chosen to be $\mu_0/2 \sim 2$~\GeV ~\cite{Bruno:2017gxd}: the high-energy one is the range $\mu_0/2 \lessapprox \mu\lessapprox \MW$ and the low-energy one is the range $\lQCD \lessapprox \mu \lessapprox  \mu_0/2 $. \cite{DallaBrida:2016uha, DallaBrida:2016kgh}
The main difference between the two is the definition adopted for the renormalised coupling $\gbar$: in the high-energy range it is the non-perturbative SF coupling \cite{Luscher:1992an}, whereas in the low-energy one it is the gradient flow (GF) coupling \cite{Fritzsch:2013je}. Note that the renormalisation condition for the quark mass renormalisation factor $1/\ZP$ is the same for both regimes, so $\sigmaP$ is expected to be a continuous function of the scale $\mu=1/L$ in the whole energy range $[\lQCD, \MW]$. 
Also note that two different gauge actions are employed in the two sectors: Wilson-plaquette action for the SF region and tree-level Symanzik improved (L\"uscher-Weisz) action for the GF one. In both cases, the fermionic action is Wilson-clover.

\section{Tuning of $\zf$ and $\hopc$}

\begin{figure}
\includegraphics[width=0.5\textwidth]{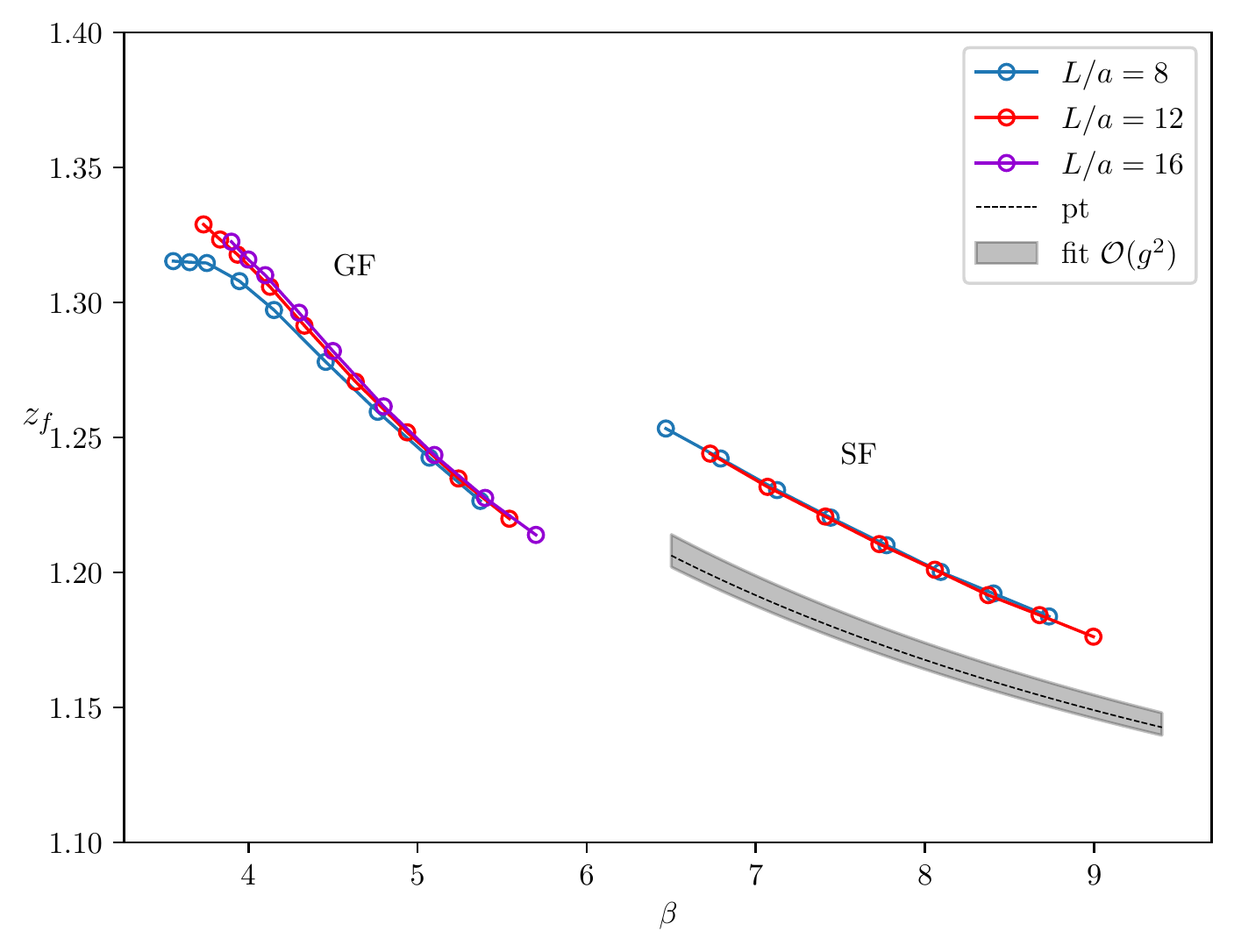}\, \, \, \, \, \, \includegraphics[width=0.5\textwidth]{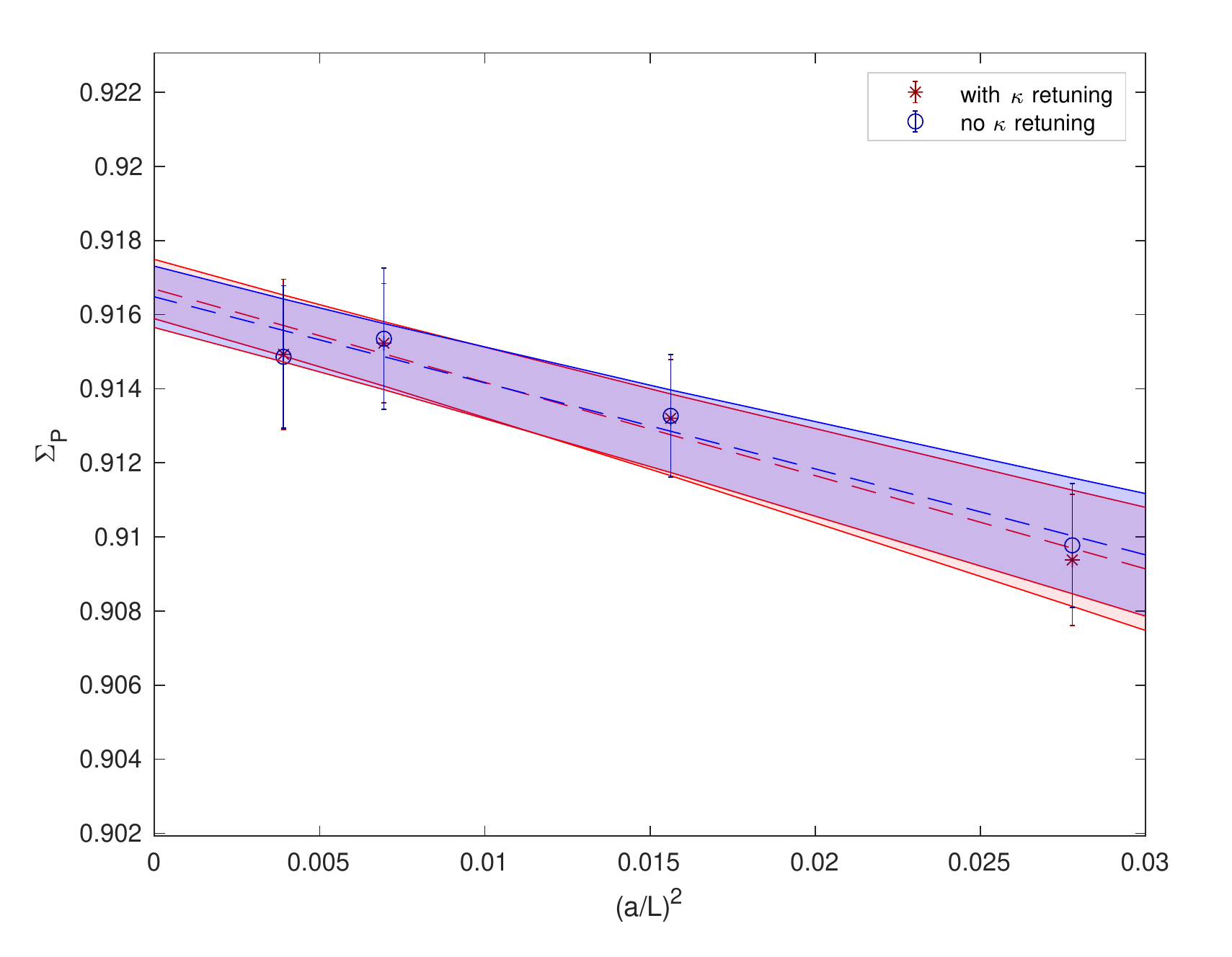}
\caption{On the left: tuning of the boundary counterterm $\zf$ on SF and GF ensembles. The black dotted line is the
perturbative result known at $O(g_0^2)$; the grey band is the
result from fitting the SF data, truncated at $O(g_0^2)$. The curve is discontinuous in $\beta$ as the definitions of the coupling and the gauge actions are different in the GF and SF regimes. \\
On the right: $\SigmaPchiSF$ vs $(a/L)^2$ at $\uSF$=2.012, with and without the retuning of $\hopc$.
\label{fig:tuning_zf_kappa}}
\end{figure}

On the lattice, the $\chi$SF version of parity (${\cal P}_5$) is broken, cf. ref.~\cite{Sint:2010eh}. We need to restore it by introducing the renormalisation coefficient $\zf$, which is tuned imposing the vanishing of a ${\cal P}_5$-odd correlation function:
\begin{equation}
\label{eq:tune-zf}
\gA^{ud}(x_0) \Big \vert_{x_0 = T/2} \,\, = \,\, 0 \,\, ,
\end{equation}
where $\gA^{ud}$ is defined analogously to $\gP^{ud}$ in ~\eqref{eq:bound_to_bulk} with an axial current in the bulk.
The tuning of $\zf$ for each ensemble is shown on the left panel of Figure \ref{fig:tuning_zf_kappa}.
\\
Moreover the hopping parameter $\kappa$ must be tuned to its critical value $\hopc$, since $\chi$SF is a mass independent renormalisation scheme. We use the SF $\hopc$ estimate of ref.~\cite{Campos:2018ahf}, obtained by imposing the vanishing of the SF-PCAC mass, $m^{\rm\scriptscriptstyle SF}(g_0^2,\kappa) \, \equiv \, \tilde{\partial}_0 \fA^{I}(x_0)/2 \fP(x_0)$\footnote{The axial current insertion in $\fA^{\scriptscriptstyle I}$ includes the $\icA$ Symanzik term.} and then tune $\zf$ in order to satisfy Eq.\eqref{eq:tune-zf}. This avoids having to retune $\kappa$ by imposing the vanishing of the  $\chi$SF-PCAC mass, $m^{\rm\scriptscriptstyle \chi SF}(g_0^2,\kappa) \, \equiv \,   \tilde{\partial}_0 \gA^{ud}(x_0)/2   \gP^{ud}(x_0)$, as proposed in ref.~\cite{DallaBrida:2018tpn}. This choice is possible because the tuning of $\kappa^{\chi \rm{SF}}$ and $\zf$ have been shown to be weakly dependent on each other ~\cite{DallaBrida:2018tpn}, suggesting that using $\hopc ^{\rm{SF}}$ in place of $\hopc^{\chi \rm{SF}}$ should not mistune $\zf$. To check this explicitly we compute the step scaling function of the pseudoscalar operator at the switching scale with two $(\hopc,\zf)$ estimates: first we use $\hopc$ of ~\cite{Campos:2018ahf} and tune $\zf$ so that Eq.\eqref{eq:tune-zf} is satisfied. Second we retune recursively $\zf$ and $\hopc$ until both condition \eqref{eq:tune-zf} and $m^{\rm\scriptscriptstyle \chi SF}=0$ are satisfied. We see in Figure \ref{fig:tuning_zf_kappa} (right panel) that  the two sets of data are perfectly compatible, both at finite lattice spacing and in the continuum.

\section{Running of the quark mass}
We compute the running of the quark mass from hadronic scales to high-energy scales and we compare our results to those of ~\cite{Campos:2018ahf}:
\begin{equation}
\label{eq:running}
\frac{M}{\mbar(\mu_{had})}=\frac{M}{\mbar(2^{k}\mu_{0})}\frac{\mbar(2^{k}\mu_{0})}{\mbar(\mu_{0}/2)}\frac{\mbar(\mu_{0}/2)}{\mbar(\mu_{had})} \ .
\end{equation}
As in ~\cite{Campos:2018ahf}, nonperturbative running factorizes into a low-energy factor computed in the GF regime (rightmost term of Eq.~\eqref{eq:running}) and a high-energy factor computed in the SF regime (central term of Eq.~\eqref{eq:running}). 
The leftmost factor is calculated by integrating the perturbative RGE at high energies (large $k$ value):
\begin{equation}
\label{eq:pt_term}
\frac{M}{\mbar(2^k \mu_0)} = 
[2 b_0 \gbar^2_{\SF}(2^k \mu_0)]^{-d_0/2 b_0} \times 
\exp \Bigl\{ -\int_0^{\gbar_{\SF}(2^k \mu_0)} dx 
\Bigl[\frac{\tau(x)}{\beta(x)} - \frac{d_0}{b_0 x} \Bigr]
\Bigr\} \,.
\end{equation}

\begin{figure}
\includegraphics[width=0.45\textwidth]{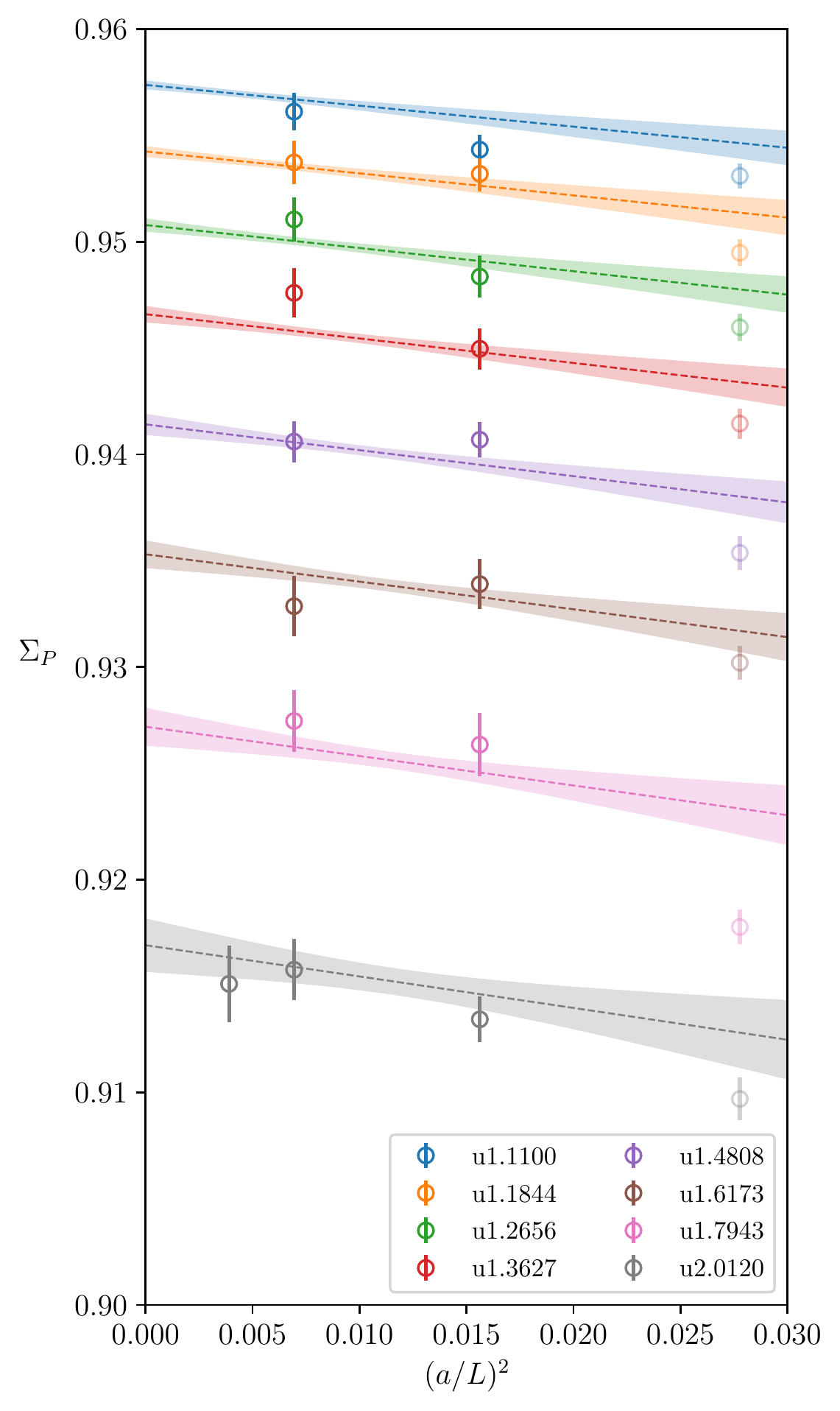} \, \, \, \, \, \, \includegraphics[width=0.45\textwidth]{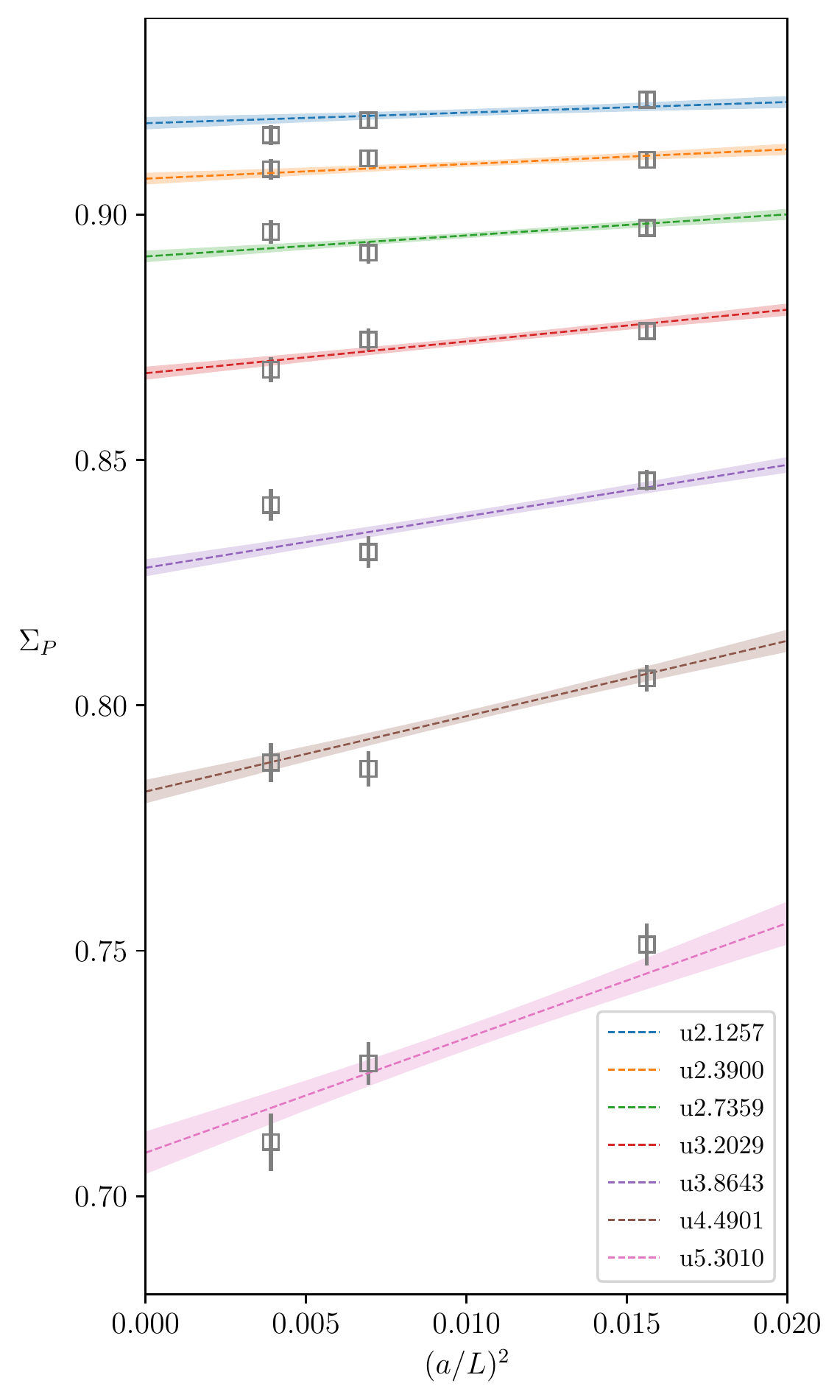}
\caption{Left: fit to the step-scaling data $\SigmaPchiSF(u, a/L)$ in the high-energy
regime; the transparent $L/a = 6$ data points are not included in the fit. The data points and the bands of the same colour are at a fixed value of the renormalised squared coupling $u$. Right:  Fit to the step-scaling data $\SigmaP$ in the low-energy regime. The bands of the same colour are at a fixed value of $u$. The data points close to the bands are at {\it approximately} the same $u$.
\label{fig:SigmaP}}
\end{figure}

\begin{figure}
\includegraphics[width=0.6\textwidth]{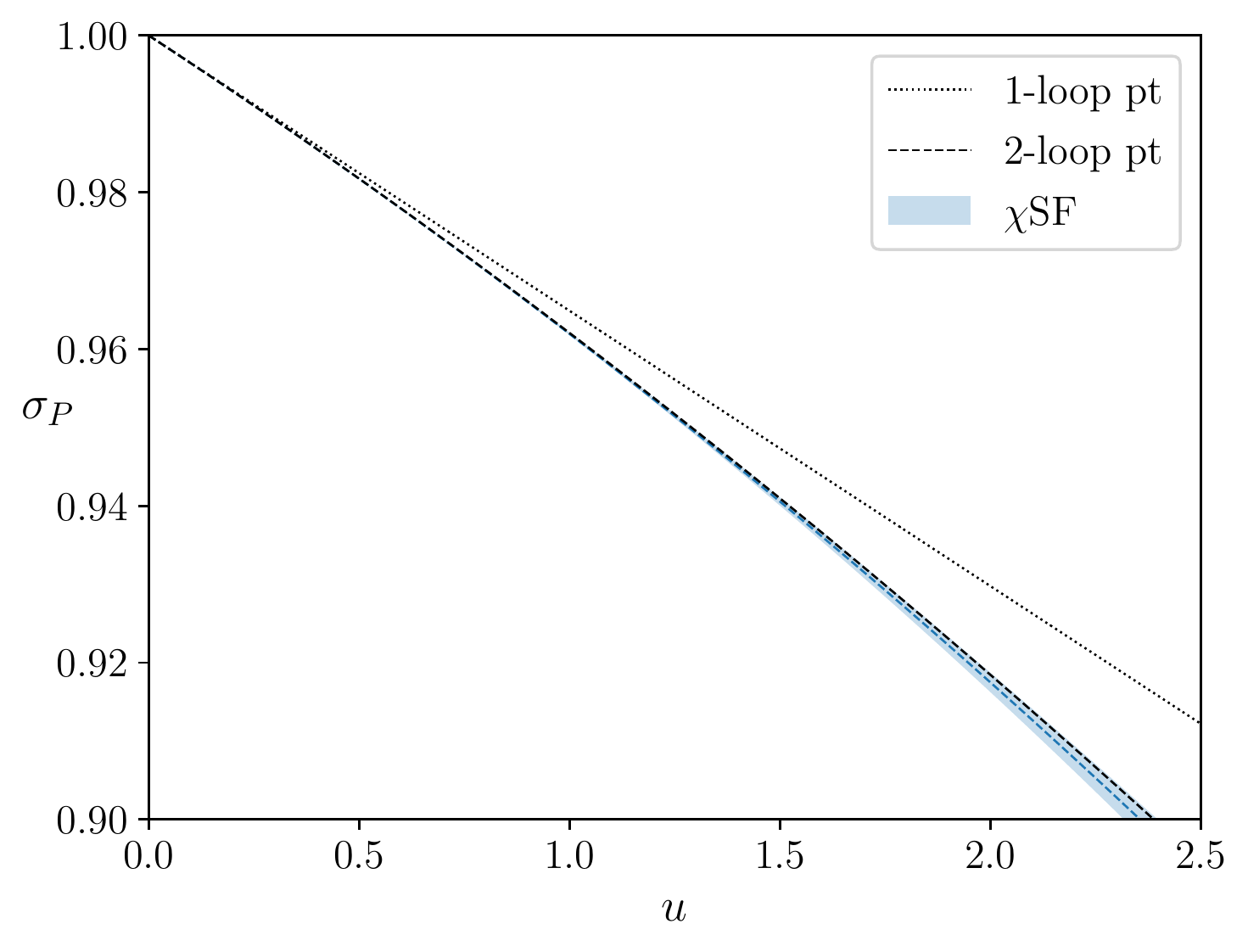}
\caption{$\sigmaP(u)$ compared with perturbation theory. 
\label{fig:sigmaP_vs_u}}
\end{figure}

To compute the central factor of Eq.~\eqref{eq:running}, the starting point is to extract the continuum $\sigmaP(u)$ from the $\SigmaPchiSF(u,a/L)$ data (see left panel of Figure \ref{fig:SigmaP})  through a fit constrained by the 1- and 2-loop perturbative coefficients of the continuum step scaling function.
In Figure  \ref{fig:sigmaP_vs_u} we show our result for $\sigmaP(u)$ and compare it to the perturbative predictions. We see that there is a very good agreement with the 2-loop perturbative result, specific to both $\chi$SF and SF schemes. 
The running of quark masses from the highest energy to the switching scale is computed in terms of $\sigmaP$: 
\begin{equation}
\label{eq:Rk}
 \frac{\mbar(2^k \mu_0)}{\mbar(\mu_0/2)} = 
\prod_{n=0}^{k} \sigmaP(u_n) \,,
\end{equation}
        where the ratio between $\mu (u_{n+1})$  and $\mu (u_n)$ is 2, as follows from the definition of the step scaling function ~\eqref{eq:sigma_def}.Having established that the result shown in Fig. \ref{fig:sigmaP_vs_u} is a robust non-perturbative estimate of $\sigmaP(u)$, even for squared couplings below the lowest simulated value $\uSF$, it is possible to compute ~\eqref{eq:Rk} even for energy scales that lie outside our simulation range, in a region where perturbation theory can be safely applied. We have checked that our results are stable for $k \ge 5$. Our preliminary result is given for the conservative choice of $k=10$. The first two factors of ~\eqref{eq:running}, i.e. ~\eqref{eq:pt_term} and ~\eqref{eq:Rk}, are combined to compute the running down to the switching scale, for which we find the following value:
\begin{equation}
\frac{M}{\mbar(\mu_0/2)}= 1.7519(74) \,.
\end{equation}
Alternatively we fit the $\SigmaPchiSF(u,a/L)$ data according to the prescription $\tau$:{\it global} of ref.~\cite{Campos:2018ahf} in order to obtain directly an estimate of the anomalous dimension $\tau(u)$ in the high-energy range. Our $\tau(u)$ result and $\beta(u)$ from ref.~\cite{Campos:2018ahf} are fed into Eq. \eqref{eq:pt_term}, written for the scale $\mu_0/2$ instead of $2^k  \mu_0$, giving $M/\mbar(\mu_0/2) = 1.7516(81)$. In a purely SF setup, ref.~\cite{Campos:2018ahf} quotes $M/\mbar(\mu_0/2) = 1.7505(89)$. All three results are in excellent agreement.

In the GF regime we fit the $\SigmaPchiSF(u,a/L)$ data (see right side of Figure \ref{fig:SigmaP}), but we must consider that the lowest bound of the GF sector is not obtained dividing the switching scale by a power of 2. Thus we prefer to compute the running in terms of the mass anomalous dimension $\tau(\gbar)$, using the relation
\begin{equation} \label{sigma_P}
\sigma_P(u) = \exp{\Bigl[
-\int_{\sqrt{u}}^{\sqrt{\sigma(u)}} dg \frac{\tau(g)}{\beta(g)}
\Bigr]} \,.
\end{equation}
Our result for the running factor between the switching and the hadronic scales
\begin{equation} 
\label{eq:R_had}
\frac{\mbar(\mu_0/2)}{\mbar(\mu_{\text{had}})} =  \exp{\Bigl[
-\int_{\sqrt{\sigma(u_0)}}^{\sqrt{u_{had}}} dg \frac{\tau(g)}{\beta(g)}
\Bigr]} \, ,
\end{equation}
is
\begin{equation}
\label{eq:val_Rhad}
\frac{{\mbar(\mu_0/2)}}{\mbar(\mu_{\text{had}})}= 0.5199(39) \, .
\end{equation}
This is also in good agreement with the result of ~\cite{Campos:2018ahf}: $\mbar(\mu_0/2)/\mbar(\mu_{\text{had}}) = 0.5226(43)$. 
Putting everything together gives:
\begin{equation}
\label{eq:total_running}
\frac{M}{\mbar(\mu_{\text{had}})} = 0.9108(78). 
\end{equation}
As expected from previous results, it agrees well with its SF counterpart of ref.~\cite{Campos:2018ahf}: $M/\mbar(\mu_{\text{had}})=0.9148(88)$. 

\acknowledgments

We wish to thank Patrick Fritzsch, Carlos Pena, David Preti, and Alberto Ramos for their help. 
This work  is partially supported by INFN and CINECA, as part of research project of the QCDLAT INFN-initiative. 
We acknowledge the Santander Supercomputacion support group at the University of Cantabria which provided access to the Altamira Supercomputer at the Institute of Physics of Cantabria (IFCA-CSIC).
We also acknowledge support by the Poznan Supercomputing and Networking Center (PSNC) under the project with grant number 466. AL acknowledges support by the U.S.\ Department of Energy under grant number DE-SC0015655.

\end{document}